\documentclass[10pt,conference]{IEEEtran}
\usepackage[square,sort,comma,numbers]{natbib} 
\usepackage{listings}
\usepackage{subcaption}
\usepackage{graphicx}

\usepackage{amsmath}

\usepackage{xcolor}

\definecolor{diffstart}{RGB}{211,211,211}
\definecolor{diffincl}{RGB}{102,205,170}
\definecolor{diffrem}{RGB}{255,69,0}
\definecolor{bostonuniversityred}{rgb}{0.8, 0.0, 0.0}
\definecolor{darkgreen}{rgb}{0.0, 0.2, 0.13}
\definecolor{yellow-green}{rgb}{0.6, 0.8, 0.2}
\definecolor{guppiegreen}{rgb}{0.0, 1.0, 0.5}
\definecolor{pastelmagenta}{rgb}{0.96, 0.6, 0.76}
\definecolor{tearose(rose)}{rgb}{0.96, 0.76, 0.76}

\usepackage{listings}
\lstdefinelanguage{diff}{
	basicstyle=\ttfamily\small,
	morecomment=[f][\color{diffstart}]{@@},
	morecomment=[f][\color{diffincl}]{+\ },
	morecomment=[f][\color{diffrem}]{-\ },
}

\makeatletter
\newif\if@restonecol
\makeatother

\usepackage[linesnumbered,ruled,vlined]{algorithm2e}
\usepackage{algpseudocode,algorithmicx}
\usepackage{amsmath}

\definecolor{dkgreen1}{rgb}{0,0.6,0}
\definecolor{gray1}{rgb}{0.5,0.5,0.5}
\definecolor{mauve1}{rgb}{0.58,0,0.82}
\usepackage{listings}
\usepackage{booktabs}
\usepackage{verbatim}
\usepackage{stfloats}
\usepackage{textcomp}
\usepackage{booktabs, tabularx}
\usepackage{rotating}
\usepackage{xcolor}
\usepackage{cite}

\usepackage{fancybox}
 \usepackage{multirow}

\usepackage{hyperref}

\usepackage{url}
\usepackage{hyperref}

\usepackage{caption}
\usepackage{subcaption}

\usepackage{balance}
\usepackage{footnote}

\usepackage{array}

\usepackage{adjustbox}
\usepackage{threeparttable}

\definecolor{dkgreen}{rgb}{0,0.6,0}
\definecolor{gray}{rgb}{0.5,0.5,0.5}
\definecolor{mauve}{rgb}{0.58,0,0.82}

\usepackage{array}
\newcolumntype{$}{>{\global\let\currentrowstyle\relax}}
\newcolumntype{^}{>{\currentrowstyle}}

\newcommand{\rqbox}[1]{
\begin{center}
\vspace{-0.2cm}
\cornersize{.1} 
\setlength{\fboxsep}{7pt}
\ovalbox{\begin{minipage}{3.3in}
{\em #1}
\end{minipage}}
\vspace{-0.2cm}

\end{center}}

\definecolor{pgrey}{rgb}{0.46,0.45,0.48}

\AtBeginDocument{%
  \providecommand\BibTeX{{%
    \normalfont B\kern-0.5em{\scshape i\kern-0.25em b}\kern-0.8em\TeX}}}

\begin{document}

\title{A Better Approach to Track the Evolution of Static Code Warnings}

\author{
	\IEEEauthorblockN{Junjie Li}
	\IEEEauthorblockA{
		Concordia University\\
		Montreal, Canada\\
		Email: l\_unjie@encs.concordia.ca}
}

\maketitle
\begin{abstract}
	Static bug detection tools help developers detect code problems. However, it is known that they remain underutilized due to various reasons. Recent advances to incorporate static bug detectors in modern software development workflows can better motivate developers to fix the reported warnings on the fly. 
	
	In this paper, we study the effectiveness of the state-of-the-art (SOA) solution in tracking warnings by static bug detectors and propose a better solution based on our analysis of the insufficiencies of the SOA solution. In particular, we examined four large-scale open-source systems and crafted a data set of 3,452 static code warnings by two static bug detectors. We manually uncover the ground-truth evolution status of the selected warnings: persistent, resolved, or newly-introduced. Moreover, upon manual analysis, we identified the critical reasons behind the insufficiencies of the SOA matching algorithm. Finally, we propose a better approach to improve the tracking of static warnings over software development history. Our evaluation shows that our proposed approach provides a significant improvement in the precision of the tracking, i.e., from 66.9\% to 90.0\%.
\end{abstract}

\begin{IEEEkeywords}
	Tranking static code warnings, Empirical study, Static analysis
\end{IEEEkeywords}






\IEEEpeerreviewmaketitle

\section{Introduction}

Static bug detection tools have been widely applied in practice to detect potential defects in software. However, they are known to be underutilized due to various reasons. First, static bug detectors detect an overwhelming number of warnings, which may be far beyond what resources are allowed to resolve~\citep{not_use_static},~\citep{beller}. Second, static bug detectors are known to detect many false positive warnings. The existence of a large number of false positives discourages developers from actively working on resolving the reported warnings~\citep{defect_types},~\citep{bugs_we_can_find}. As a result, a significant portion of static code warnings remain unresolved by developers and can hinder software quality. Researchers have been working on techniques to improve the performance of static bug detectors~\citep{quinn, fix_warning_first}.

Recent studies show that by integrating static bug detectors in software development workflows, such as code review, developers demonstrated a higher response rate in resolving the reported static warnings~\citep{google_static, tricoder}. Tracking the evolution of static code warnings reveals which warnings are \textit{persistent}, i.e., unresolved by developers, which warnings are \textit{resolved}, and which warnings are \textit{newly introduced} by the latest changes. Currently, there has been little effort to review the existing solutions to track the evolution of static code warnings. Avgustinov et al.~\citep{Tracking15} presented the first algorithm that combines various information of one warning. It compares two warnings in layers and eventually establishes mappings between two sets of warnings using three matching strategies(i.e., \texttt{Location Matching}, \texttt{Snippet Matching} and \texttt{Hash Matching}), which we refer to as the state-of-the-art (SOA) solution. The information of one warning they used is shown as Figure \ref{fig:warningInstance}. However, they did not present an evaluation of the performance of the SOA solution in tracking the static code warnings. In other words, it remains unknown whether or not the SOA solution has an acceptable performance.  

\begin{figure}[h]
	\lstset{
		language=XML,
		morekeywords={encoding,
			WarningType,WarningInstance,Commit,Class,Method,Field,Project,FilePath,StartLine,EndLine,WarningInstance, WarningType}
	}
	\begin{lstlisting}
	<WarningInstance> 
	<WarningType>SE_BAD_FIELD</WarningType>
	<Project>jclouds</Project> 
	<Class>ContextBuilderTest</Class>
	<Method></Method>
	<Field></Field>
	<FilePath>org/jclouds/ContextBuilder.java</FilePath>
	<StartLine>70</StartLine>
	<EndLine>75</EndLine> 
	</WarningInstance>
	\end{lstlisting}
	\caption{An example of the representation of one static code warning. The representation has been simplified to only show the information used by the SOA matching approach.}
	\label{fig:warningInstance}
\end{figure}

In this paper, we collected and uncovered the ground-truth evolution status between two consecutive commits using two widely used static bug detectors(\textit{PMD}~\citep{PMD19} and \textit{Spotbugs}~\citep{Spotbugs19}) on two open-source software projects (\textit{JClouds} and \textit{Kafka}). We examined the SOA solution in tracking the evolution of static code warnings. Our investigation shows that the SOA solution achieves inadequate results, and we performed a manual analysis to uncover the insufficiencies of the SOA solution. Based on our findings, we proposed a better approach to track the static code warnings. The evaluation based on the crafted data set shows that our approach can significantly improve the tracking precision. 	 

\section{uncovering Ground-truth and approach improvement}
\label{sec:study}
\noindent\textbf{Static Bug Detectors and Analyzed Open-source Systems.} In this paper, we include two static bug detectors, i.e., \textit{PMD} and \textit{Spotbugs}. In terms of Analyzed systems, our study includes four Java open-source systems, \textit{JClouds}, \textit{Kafka}, \textit{Spring-boot} and \textit{Guava}. Two of the systems(i.e., \textit{JClouds} and \textit{Kafka}) are used to are used to uncover the ground-truth. The other two systems(i.e., \textit{Spring-boot} and \textit{Guava}) are selected to evaluate both approaches without biases.

\noindent\textbf{A Description of the SOA Approach.} For those warnings that cannot be matched by \texttt{Exact matching}(i.e., warnings from two revisions will be matched up when their metadata like Figure~\ref{fig:warningInstance} are identical), there are three matching strategies in the SOA approach. The first one is \texttt{Location matching}. It is based on the \textit{diffs}~\citep{hunt1977diff} ~\citep{myers1986diff} between two revisions. When \texttt{Location matching} fails, \texttt{Snippet matching} will be used. Given the source location defined by a start line and an end line, code snippets in between are extracted from both revisions. \texttt{Snippet matching} will decide a mapping if they are identical. The two strategies can only match warnings in the same class file. When a file is moved to a new location (i.e., file path are modified), they cannot handle it. For such case, \texttt{Hash matching} approach can be helpful. This matching strategy tries to match warnings based on the similarity of their surrounding code.

\noindent\textbf{Improved Approach.} 
We re-implemented the SOA approach and applied it on analyzed open-source systems. Overall, We crafted a dataset of 1,715 static code warnings and manually uncovered their ground-truth evolution status. In short, the overall precision of the SOA approach on the collected dataset is only 62.4\%, which means that 37.6\% \textit{resolved} or \textit{newly introduced} warnings actually are \textit{persistent}. Guided by our manual analysis results, we propose to improve the SOA approach by better handling refactoring changes. In particular, our proposed approach reuses the two matching strategies(i.e., \texttt{Location matching} and \texttt{Snippet matching}) of the SOA approach and revise a few key steps to improve the inaccurate tracking.
\begin{enumerate}
	\item Improvement 1 - Including refactoring. From our dataset, we find that code refactoring will modify the metadata(i.e., Figure~\ref{fig:warningInstance}) of static warnings, which causes that three matching strategies fail to work. To address this insufficiency, we include the refactoring information to improve the tracking using RefactoringMiner~\citep{RefactoringMiner18}. When the location of a static code warning having code refactoring is detected, our approach can take a better and correct mappings by modifying the metadata of the pre-commit warning to the post-commit warning. Then we take the two matching strategies to match them. In particular, \texttt{Hash matching} in the SOA approach is designed to handle the case of the class files renamed or moved that are included into refactoring information. Thus we remove \texttt{Hash matching}.  
	\item Improvement 2 - Adopting \textbf{Hungarian algorithm}.  When a pair of warnings is mismatched from the SOA approach, the others will be affected. We do find such false positives due to the SOA mismatching. 
	To address this insufficiency, we adopt the \textbf{Hungarian algorithm}~\citep{kuhn1955hungarian}, a classic approach to solve the assignment problem in bipartite graphs, to reduce the effect of the matching strategies order. When a pair of candidates is found in two matching strategies(i.e., \texttt{Location Matching} and \texttt{Snippet Matching}), instead of deciding it as a matched pair directly in the SOA approach, we construct a Hungarian matrix to save matching candidate pairs. After saving all matching candidate pairs, we leverage maximum matching on them to decide the matched pairs.
\end{enumerate}

\section{An Evaluation of our approach}
\label{sec:evaluation}
We propose an improved approach based on the manual analysis on {\textit JClouds} and {\textit Kafka}. To avoid a biased evaluation, in addition to the two systems, we also select two other open-source software systems (i.e., {\textit Spring-boot} and {\textit Guava}) to evaluate our improved approach and show how much improvement our approach has compared to the SOA approach. Totally, we collect 3,452 static warnings from SOA approach. 

Since tracking the static code warnings is not a standalone task for each individual warning, it is, in fact, a mapping problem between two sets. Hence, we actually applied our improved approach on all static warning on a commit, which is a superset of the 3,452 warnings in the manually-labeled dataset. The remaining warnings while not in our crafted dataset have a pre-assumed satuts, ``\textit{persistent}''.

\begin{table}[h]
\caption{The performance comparison between the SOA approach and our approach. Note that FP is short for false positive. A lower FP ratio is desired.}
\label{tab:evaluation}
\centering
\scalebox{0.74}{
	
\begin{tabular}{crrrr}
\toprule
& \multicolumn{2}{c}{Resolved} & \multicolumn{2}{c}{ Newly-Introduced} \\\cline{2-5}
& FP (SOA) & FP (our approach)  & FP (SOA) & FP (our approach)\\
\textbf{PMD}  &&&& \\\hline
JClouds & 63.6\% (178/280) & 3.8\% (4/106) & 36.8\% (57/155) & 8.4\% (9/107)\\ 
Kafka & 45.4\% (148/326)  & 27.0\% (66/244) & 8.6\% (22/255) & 4.1\% (10/243) \\\
Spring-boot & 36.7\% (80/218) & 13.8\% (22/160) & 42.3\% (80/189) & 16.8\% (22/131)\\ 
Guava & 41.9\% (124/296)  & 8.0\% (15/187) & 66.0\% (124/188) & 19.0\% (15/79) \\\hline
\textbf{Spotbugs}  &&&& \\\hline
Jclouds & 17.3\% (18/104) & 1.1\% (1/87) & 17.9\% (14/78) & 3.0\% (2/66) \\
Kafka & 37.8\% (114/301) & 15.8\% (35/222) & 43.5\% (94/216) & 17.0\% (25/147) \\\
Spring-boot & 3.6\% (7/193) & 0.5\% (1/187) & 4.4\% (7/160) & 0.6\% (1/154)\\ 
Guava & 20.1\% (58/289)  & 6.1\% (15/246) & 26.0\% (53/204) & 6.2\% (10/161) \\\hline
\textbf{Total}  & 36.2\% (727/2007) & 11.1\% (159/1437) & 31.2\% 451/1445) &  8.6\% (94/1087)\\
\bottomrule
\end{tabular}
}
\end{table}

Table~\ref{tab:evaluation} lists the comparison results between the SOA approach and our improved approach. In the 3,452 warnings, the SOA approach has 1,178 warnings with a wrong evolution status. Compared to that, our proposed approach reduces the false positives significantly, from 1,178 to 253, i.e., the false positive rate drops to 10.0\%. Our approach maps correctly for the labeled persistent warnings, which are mistakely labeled as \textit{resolved} or \textit{newly-introduce} by the SOA approach.

\section{Conclusions}
\label{sec:conclusions}
Tracking the evolution of static code warnings across software development history becomes a vital question due to the increasing interest to further utilize static bug detectors by integrating them in developers' workflow. Also, such tracking is widely used in many downstream software engineering tasks. This study presents an investigation on the performance of the state-of-the-art approach in tracking static code warnings. In particular, a dataset of 3,452 static code warnings and their evolution status is crafted.
Last, this paper present an improved approach, which is shown to outperform the SOA approach significantly in terms of the tracking precision.

\bibliographystyle{ieeetran}
\bibliography{paper}

\end{document}